\documentclass[12pt]{article} 
\usepackage{amsmath}
\usepackage{latexsym}
\usepackage{times}
\begin{document} 
\bibliographystyle{unsrt} 
 
%\vbox{\vspace{38mm}} 
\begin{center} 
{\bf Chiral Symmetry and Vertex Symmetry in the extended Moeller-Rosenfeld Model}\\[20mm] 
 
Sultan Catto$^{\dag}$\\{\it Physics Department\\ The Graduate School and University Center\\365 Fifth Avenue\\
New York, NY 10016-4309\\ and \\ Center for Theoretical Physics \\The Rockefeller University \\
1230 York Avenue\\ New York NY 10021-6399}\\[10mm] 
\end{center} 
\vbox{\vspace{3mm}}

\begin{abstract} 
Vertex symmetry for interacting fermions will be shown to lead to 
a Lagrangian exhibiting $SU(2N)_{W}$ invariance associated with the subgroup 
$SU(2N)_{q} \times SU(2N)_{\bar{q}}$ generated by C-odd and C-even spin operators.
Approximate $SU(6)_W$ vertex symmetry as well as chiral invariance will then be shown to follow from a principle of maximum smoothness (M\"oller-Rosenfeld) of the bound state quark wave function.
\end{abstract}
\vbox{\vspace{10mm}}


\vbox{\vspace{4mm}}

$^\dag$ Work supported in part by DOE contracts No. DE-AC-0276 ER 03074 and 03075, and PSC-CUNY Research Awards.
\newpage
\section*{Introduction}

In this paper we shall connect some approximate 
symmetries of nuclear forces to what is believed to be the fundamental 
theory of nuclear forces: Quantum Chromodynamics (QCD). Ideally we should be
able to calculate all hadronic constants like the meson-baryon, meson-meson
coupling constants (more generally the invariant form factors) and baryon and
meson masses to give the strength and the range of nuclear forces, starting
from the dynamics of quarks and gluons. We are still far from this goal. 
However, we can set ourselves more modest objectives. As an intermediate step
we might concentrate on the properties of the coupling of mesons to quarks,
then try to infer general properties of hadronic interactions from the
assumed properties of the former. If we study the nucleon-nucleon interaction,
say from the nucleon-nucleon, meson-nucleon scattering, or the spectra of
light nuclei, two kinds of approximate symmetries emerge: (a) A Wigner $SU(4)$
symmetry combining isospin with spin and broken by spin dependent terms; (b) An
approximate chiral symmetry broken by the pion mass. Recently more evidence
has accumulated in nuclear physics extending the domain of validity to more 
hadronic processes$^{\cite{mosh}}$. The important role played by the $\Delta$ 
exchange along nucleon exchange in phenomena exhibiting chiral symmetry points 
out to a possible link of the $SU(4)$ symmetry (which puts $N$ and $\Delta$ 
in the same multiplet) with chiral symmetry$^{\cite{cats}}$. In turn this relation can only 
arise from the combination of the $SU(4)$ and chiral symmetries at the deeper 
quark level as a consequence of QCD. Another piece of evidence comes from the
role the scalar mesons $\epsilon$(f) and $\delta$($a_{0}$) play in decay 
processes. Scalar meson couplings calculated by this method will be shown 
to be consistent with chiral and $SU(4)$ symmetries at the quark level. This
fundamental $SU(4) \times SU(4)$ symmetry leads to a universal coupling of 
quarks to pseudoscalar, vector mesons as well as to scalar and pseudovector 
mesons. In turn the exchange of all these mesons between nucleons yield 
nuclear forces exhibiting an approximate chiral $SU(4) \times SU(4)$ symmetry.
Arguments supporting this view will be given in this paper. It will be shown 
that for relativistic quarks the potential is invariant under chiral 
transformations and the $SU(4)$ group except for definite spin dependent terms
and mass differences between mesons.
     
Starting from field theory of free and interacting quarks we will first examine quark spin and solve a quantum mechanical bound state problem as an approximation to field theory bound state problem by extracting a potential $V$. We will then study the general properties of the confining (long range) part of $V$ and its short range part and address models leading to a spin independent or $SU(N\times 2)$ invariant potentials (where $N$ is the number of flavors). 
For relativistic quarks we will show that the potential is invariant under chiral transformations 
and the $SU(4)$ group except for definite spin dependent terms and mass 
differences between mesons, and show invariance of the potential using a larger group structure.

\section*{Quark-meson couplings}
The fundamental couplings in QCD are the self coupling of the gluons and the
gluon-quark couplings which are determined uniquely by the gauge principle of 
local color invariance $(SU(3)^{c})$. Through this interaction quarks are 
assumed to form $q \bar{q}$ (meson) and $(qqq)$ (baryon) bound states. Because
of asymptotic freedom, quarks behave like quasi-free particles at small 
separation. There is a vertex function between the $q \bar{q}$ meson bound
state and the quasi free $q$ and $\bar{q}$. For each quantum state of the 
$q \bar{q}$ system this vertex function can be approximated by a constant which
we shall call the quark-meson coupling constant $g_{q \bar{q} M}$. Then, under 
the additivity hypothesis the meson-nucleon coupling constant is

\begin{equation}
\Gamma_{N \bar{N} M} = 3~ g_{q \bar{q} M} .
\end{equation}
The relativistic coupling constants are defined in terms of an effective 
quark meson interaction Lagrangian

\begin{eqnarray}
{\cal L}_{int} & = & g_{s}~\bar{q} \tau_{a} q \chi^{a} + h_{s}~\bar{q} \tau_{a}
\sigma_{\mu \nu} q \partial_{\mu} \chi^{a} +  
g_{Psv}~\bar{q} \tau_{a} \gamma_{5} \gamma_{\mu} q W_{5 \mu}^{a} \nonumber \\ & & +
h_{Psv}~ \bar{q} \tau_{a} \sigma_{\mu \nu} q \tilde{W}_{\mu \nu}^{a} 
+ g_{Ps}~\bar{q} \gamma_{5} \tau_{a} q \phi^{a} + h_{Ps}~ \bar{q} \tau_{a} 
\gamma{_5} \gamma_{\mu} q \partial_{\mu} \phi^{a}  \nonumber \\ && +   
g_{v}~\bar{q} \tau_{a} \gamma_{\mu} q V_{\mu}^{a} + h_{v}~ \bar{q} \tau_{a} 
\sigma_{\mu \nu} q V_{\mu \nu}^{a}                \label{eq:pa}
\end{eqnarray}
where $a = 0, 1, 2, 3$, $\tau_{0} = 1$ and

\begin{equation}
V_{\mu \nu}^{a} = \partial_{\mu} V_{\nu}^{a} - \partial_{\nu} V_{\mu}^{a} ,~~~~ 
W_{\mu \nu}^{a} = \partial_{\mu} W_{5 \nu}^{a} -\partial_{\nu} W_{5 \mu}^{a} , 
\end{equation}

\begin{equation}
\tilde{W}_{\mu \nu}^{a} = \frac{1}{2} \epsilon_{\mu \nu \rho \kappa} 
W_{\rho \kappa}^{a} ,   
\end{equation}

\begin{equation}
 \partial_{\mu} V_{\mu}^{a} = 0,~~~~~~ \partial_{\mu} W_{5 \mu}^{a} = 0 .
\end{equation}
Also note that the $h_{s}$ term in Eq.(\ref{eq:pa}) is a total divergence
and can be omitted. The derivative (momentum dependent) interactions are
important since ${\cal L}_{int}$ is an effective Lagrangian and there is no 
renormalizability constraint in the Lagrangian. The constants $g$ and $h$ can
be calculated from QCD following a definite scheme. Forexample at short 
distances coupling is similar to the probability of dissociation of positronium
into its constituents, the massless photon field playing the role of the gluon
field. For group theoretical treatment of this problem see Barut and Aydin$^{\cite{baru}}$. Rather than embarking on such a calculation we can review the available phenomenological information on these couplings, starting from the nucleon-meson vertex.

Also note that because $SU(4) \times SU(4)$ is embedded in $SU(6) \times SU(6)$,
there are two pseudoscalar fields $\phi^{0}$, namely $\eta$ and $\eta^{'}$ and
two vector fields $V_{\mu}^{0}$, namely $\omega_{\mu}$ and $\phi_{\mu}$, with
$\eta^{'}$ being approximately $SU(4)$ singlet. There is a similar doubling of
isosinglets for scalars and pseudovectors. We have the following 
identifications:
 
\begin{equation}
\begin{array}{cc}
\phi^{0}: & \eta, \eta^{'}   \\
\phi^{i}: & ~~~~~~~\pi^{i}~~~(i=1, 2, 3)   \\
V_{\mu}^{0}: & \omega_{\mu}, \phi_{\mu}  \\
V_{\mu}^{i}: & \rho_{\mu}^{i}  \\
\chi^{0}: & \epsilon (f_{0})  \\
\chi^{i}: &  \delta^{i} (a_{0}^{i})   \\
W_{\mu}^{0}: & E_{\mu} \\
W_{\mu}^{i}: & ~~~~A_{\mu}^{(1) i} (a_{\mu}^{(1) i})
\end{array}
\end{equation}

To determine the representation we start from the quarks $q_{L}^{\alpha}$ and
$q_{R}^{\alpha}$ ($\alpha = 1, 2$), with $\alpha$ the isospin index, and with further identifications

\begin{eqnarray}
q_{L}^{\alpha} ~:~(4, 1),~~~~q_{R}^{\alpha}~:~(1, 4)  \nonumber \\
\hat{q}_{L}^{\alpha}~= i \sigma_{2} q_{L}^{*}:~(\bar{4}, 1),~~~~\hat{q}_{R}^{\alpha} = 
i \sigma_{2}~q_{R}^{*}~:~ (1, \bar{4})
\end{eqnarray}

\begin{equation}
\bar{q} \tau^{a} q = q_{L}^{\dag} \tau^{a} q_{R} + q_{R}^{\dag} \tau^{a} q_{L},
~~~~ \bar{q}  \gamma_{5} \tau^{a} q =
q_{L}^{\dag} \tau^{a} q_{R} - q_{R}^{\dag} \tau^{a} q_{L} 
\end{equation}
so that

\begin{eqnarray}
\bar{q} \frac{1 + \gamma_{5}}{2} \tau^{a} q \longleftrightarrow
q_{L}^{\dag} \tau^{a} q_{R} ~~:~(\bar{4}, 4)~~~~~~\mbox{\boldmath $\pi$},~~~~~
\mbox{\boldmath $\delta$} ({\bf a}_{0})   \nonumber \\
\bar{q} \frac{1 - \gamma_{5}}{2} \tau^{a} q \longleftrightarrow
q_{R}^{\dag} \tau^{a} q_{L}~~:~(4, \bar{4})~~~~~~\epsilon~(\eta, ~f_{0})  \nonumber \\
\bar{q} \frac{1 + \gamma_{5}}{2} \tau^{a} \sigma_{\mu \nu} q \longleftrightarrow
q_{L}^{\dag} \tau^{a} \sigma^{i} q_{R}~~:~(\bar{4}, 4)~~~~~~\rho_{\mu \nu}^{a},~
B_{\mu \nu}^{a}~(A_{\mu \nu}^{a}) \nonumber \\
\bar{q} \frac{1 - \gamma_{5}}{2} \tau^{a} \sigma_{\mu \nu} q \longleftrightarrow
q_{R}^{\dag} \tau^{a} \sigma^{i} q_{L}~~:~(4, \bar{4})~~~~~~\omega_{\mu \nu},
B_{\mu \nu}^{0}~(a_{\mu \nu}^{0}) \nonumber \\
\bar{q} \tau^{a} \frac{1+\gamma_{5}}{2} \gamma_{\mu} q \longleftrightarrow
q_{L}^{\dag} \tau^{a} \sigma_{\mu} q_{L} ~~:~(15, 1) + (1, 1)  \nonumber \\
\bar{q} \tau^{a} \frac{1-\gamma_{5}}{2} \gamma_{\mu} q \longleftrightarrow
q_{R}^{\dag} \tau^{a} \sigma_{\mu} q_{R} ~~:~(1, 15) + (1, 1)
\end{eqnarray}
Thus, each meson is described by its generalized coordinates and generalized
momenta (derivative field). The combination of these fields in the interaction
Hamiltonian generate the $(4, \bar{4})$, $(\bar{4}, 4)$, $(15, 1)$, $(1, 15)$
and $(1, 1)$ representation of the chiral $SU(4) \times SU(4)$.

The Dirac $q^{\alpha}$ quark Lagrangian can be regarded as being $SU(4) \times
SU(4)$ invariant since the $u$ and $d$ quarks can be taken to have negligible 
masses in the first approximation.

A mass term for the quarks that can be induced by a non-zero vacuum 
expectation value  of the isosinglet scalar field will behave like the neutral
component of $(4, \bar{4}) + (\bar{4}, 4)$, breaking the symmetry down to 
$SU(4)$ and isospin $SU(2)$.

Under the $SU(4)$ subgroup of $SU(4) \times SU(4)$, the nucleon $N$ and the
$\Delta$ resonance form the $20$-dimensional representation contained in the
symmetric part of the product $4 \otimes 4 \otimes 4$. The various derivative
couplings of $N$ and $\Delta$ are contained in $[(4, 1) + (1, 4)]^{3}$ 
product representations of $SU(4) \times SU(4)$.

Let us now briefly look at the free quark approximation to QCD and vertex
symmetry for interacting fermions, then buildup the symmetry in the potential
approximation.  We shall later return also to embedding these structures
in larger groups and look at group decompositions.

\section*{ The Free Quark Approximation to QCD}
First consider the simplified case of colored quarks with no
flavor, interacting through the exchange of color gluons and obeying 
the Dirac equation

\begin{equation}
\gamma_{\mu} D_{\mu} \psi^{i} = m \psi^{i}~~~~~~ (i = 1, 2, 3)
\end{equation}
where m is the quark mass, $i$ the color index and $D_{\mu}$ the covariant derivative

\begin{equation}
D_{\mu} =\partial_{\mu} + \frac{i}{2} g \lambda^{a}
B_{\mu}^{a} ,~~~~~~ (a = 1, \ldots, 8)  
\end{equation}
where $g$ is the quark-gluon coupling constant and the
$\lambda$'s are the eight $3 \times 3$ Gell-Mann matrices associated 
with $SU(3)$. When we consider a quark and an antiquark localized at 
very short distance separation, then, from asymptotic freedom$^{\cite{pol}}$ 
we know that the renormalized strong fine structure constant 
${g_{r}^{2}} / 4 \pi = \alpha_{s}$ tends to zero at high momentum 
transfers (short spatial separation). In
such a situation the covariant derivative $D_{\mu}$ can be replaced by
$\partial_{\mu}$ and in the zeroth approximation we can treat the quark inside
the hadron as obeying the free Dirac equation for small $x$, the
origin being the center of mass of the quark system in the hadron. Then we can write

\begin{equation}
\psi (x) = \int \frac{d^{3} p}{\sqrt{2 p_{0}}} e^{\frac{1}{2} \gamma_{5} 
\mbox{\boldmath $\sigma$} \cdot \mbox{\boldmath $ \lambda$}(p)} e^{i \gamma_{4} \mbox{\boldmath$p$} \cdot \mbox{\boldmath$x$}} (a_{{\bf p}} + \gamma_{5} \hat{b}_{{\bf p}}) 
\end{equation}
where

\begin{equation}
a = \left(
\begin{array}{c}
a_{1} \\  a_{2}
\end{array}     \right) ,~~~~ \hat{b} = i \sigma_{2} b^{*} = \left(
\begin{array}{c}
b_{2}^{\dag} \\ - b_{1}^{\dag}
\end{array}     \right)  
\end{equation}
                                 
\begin{equation}
\mbox{\boldmath $\lambda$}(p) = \frac{{\bf p}}{|{\bf p}|} \tanh^{-1} 
\frac{|{\bf p}|}{p_{0}} 
\end{equation}

\begin{equation}
a_{p}= \left(
\begin{array}{c}
a \\  a
\end{array}     \right) = \gamma_{4} a_{p},~~~~~{\rm if}~~ \gamma_{5}~~ {\rm is~~ diagonal} \nonumber
\end{equation}
and

\begin{equation}
a_{p}= \left(
\begin{array}{c}
a \\  0
\end{array}     \right)= \gamma_{4} a_{p} ,~~~~~{\rm if}~~\gamma_{4}~~{\rm is ~~diagonal}
\end{equation}
$a_{p, s_{z}}^{\dag}, b_{p, s_{z}}^{\dag}$ ($s_{z} = 1,2$) are respectively 
creation operators for particle and antiparticle Wigner states for the 
Poincar\'{e} group corresponding to momentum ${\bf p}$ and spin component 
$s_{z}$, i.e.

\begin{equation}
 a_{p, s_{z}}^{\dag}|0> = |m, s = \frac{1}{2};{\bf p}, s_{z}>  
\end{equation}
with
\begin{equation}
p_{0}^{2} - |{\bf p}|^{2} = m^{2}
\end{equation}

The exponential can be evaluated as

\begin{equation}
exp\{ \frac{1}{2} \gamma_{5} \mbox{\boldmath $\sigma$} \cdot \mbox{\boldmath $\lambda$}(p)\} =\sqrt{
\frac{p_{0} + \gamma_{5} \mbox{\boldmath $\sigma$} \cdot {\bf p}}{m}}=
\frac{m+p_{0} + \gamma_{5} \mbox{\boldmath $\sigma$} \cdot {\bf p}}
{\sqrt{2 m (m+p_{0})}}  
\end{equation}
We also have the useful identity

\begin{equation}
exp\{ \frac{1}{2} \gamma_{5} \mbox{\boldmath $\sigma$} \cdot \mbox{\boldmath 
$\lambda$}(p)\}
\frac{1+\gamma_{4}}{2} = \sqrt{\frac{p_{0}}{m}} (\frac{m+i \mbox{\boldmath $\gamma$} 
\cdot {\bf p}}{p_{0}})^{\frac{1}{2}} \frac{1+\gamma_{4}}{2}     \label{eq:exp}
\end{equation}
The projection operators $\frac{1}{2}  (1 \pm \gamma_{4})$ project the 
large $(+)$ or small $(-)$ components of the free wave function. To 
simplify the formulae we have also omitted the color index. For a more
general description in terms of boost operators and Heisenberg algebras
satisfied by these operators we refer to our earlier paper$^{\cite{scpr}}$.

We have the following conserved quantities

(a) Quark Number:

\begin{equation}
 B = \int j_{\mu} d \sigma_{\mu} = \int j_{0} d^{3} x 
\end{equation}
where

\begin{equation}
j_{\mu} = \bar{\psi}^{i} \gamma_{\mu} \psi^{i} , ~~(i = 1, 2, 3);~~~~
\partial_{\mu} j_{\mu} = 0 
\end{equation}
This is the number of quarks minus the number of antiquarks.

(b) Charges

Putting back the flavor index $r$ in the case of $3$ quarks
\begin{equation}
Q^{a} = \frac{1}{2} \int \bar{\psi}^{r} \gamma_{\mu} 
\lambda_{rs}^{a} \psi^{s} d \sigma^{\mu} 
\end{equation}

(c) Energy Momentum 

\begin{equation}
P_{\mu} = \int T_{\mu \nu} d \sigma^{\nu}, ~~~~~~ T_{\mu \nu} =
\bar{\psi}  \gamma_{\mu} \overleftrightarrow{{\partial}_{\nu}} \psi 
\end{equation}
The Hamiltonian $P^{0}$ is given by:

\begin{equation}       
H = P^{0} = \int p^{0} (a_{p}^{\dag} a_{p} + b_{p}^{\dag} b_{p}) d^{3} p 
\end{equation}

(d) Conserved Spin Operators

We now make the Tani$^{\cite{tan}}$-Foldy-Wouthuysen$^{\cite{fw}}$ transformation
which is the same as the Wigner transformation on particle states. 
Let us rewrite the Dirac equation in the form

\begin{equation}
i H \psi = (i m \gamma_{4} - i \gamma_{4} \gamma_{n} \partial_{n}) \psi      \label{eq:hb}
\end{equation}
($i \partial_{n} = p_{n}$ on a momentum eigenstate).

     The covariant states are obtained by applying on the vacuum 
the operator $\psi^{c} = \gamma_{2} \psi^{*}$, the charge conjugate operator 
obtained by exchanging $a$'s and $b$'s. 

\begin{equation}
\psi^{c} |0> = \int d^{3} p~ exp \{ \frac{i}{2} \mbox{\boldmath $\gamma$} \cdot \mbox{\boldmath 
$\lambda$}
({\bf p}) \} |{\bf p}> ~e^{i \mbox{\boldmath $p$} \cdot \mbox{\boldmath $x$}}
\end{equation}
where we have used 

\begin{equation}                                               
\gamma_{5} \mbox{\boldmath $\sigma$} = i \gamma_{4} \mbox{\boldmath $\gamma$},~~~~
\gamma_{4} \hat{a}_{p} |0> = \hat{a}_{p} |0> 
\end{equation}
and Eq.(\ref{eq:exp}). To obtain the covariant state from the
Wigner state we must then apply a non-local operator on $|{\bf p}>~exp(ip \cdot x)$. We find

\begin{equation}
\psi^{c} |0> = W(-i \mbox{\boldmath $\nabla$} ) \int \hat{a}_{p} ~e^{i p \cdot x} (d^{3}p)|0>  
\end{equation}
where

\begin{equation}
W(-i \mbox{\boldmath $\nabla$} ) = \sqrt{\frac{m+\mbox{\boldmath $\gamma$} \cdot \mbox{\boldmath $\nabla$}}{(m^{2} - 
\nabla^{2})^{\frac{1}{2}}}} 
\end{equation}
so that the operator

\begin{equation}
\varsigma (x) = W(-i \mbox{\boldmath $\nabla$} ) \psi (x)                     
\end{equation}
creates Wigner antiquark states and $\varsigma^{c}(x)$ creates quark states.
The non-local operator $W(-i \mbox{\boldmath $\nabla$})$ is a unitary operator that connects
covariant states with Wigner states.

We can now transform the Hamiltonian with $W$. Transforming Eq.(\ref{eq:hb}) we find

\begin{equation}
H^{'}= W H W^{-1} = \gamma_{4} E_{p}= \gamma_{4} \sqrt{m^{2} +
|{\bf p}|^{2}} = \gamma_{4} (m^{2} - \nabla^{2})^{\frac{1}{2}}        \label{eq:hc}
\end{equation}

Thus, the transformation diagonalizes $H$ in a representation for
which $\gamma_{4}$ is diagonal. The expansion of Eq.(\ref{eq:hc}) in 
$m^{-2} \nabla^{2}$ gives
the Schr\"{o}dinger operator for the upper and lower components
separately, plus relativistic correction terms. Note that $H^{'}$,
unlike $H$, commutes with the spin operators $\sigma_{n}$ and with $\gamma_{4} \sigma_{n}$

\begin{equation}
[\sigma_{n}, WHW^{-1}] = 0 
\end{equation}
or

\begin{equation}
[W^{-1} \sigma_{n} W, H] = 0 
\end{equation}
Hence, the non-local objects

\begin{equation}
\frac{1}{2} \sigma_{n}^{'} = \frac{1}{2}  W^{-1} \sigma_{n} W,~~~~~
\frac{1}{2} (1 \pm \gamma_{4}) \sigma_{n}^{'}
\end{equation}
are constants of the motion. These are the non-covariant spin
operators that are conserved for free quarks and approximately
conserved in QCD. Thus the conserved generators of the relativistic 
but non-covariant $SU(6) \times SU(6)$ are

\begin{equation}
\frac{1}{2} (1 \pm \gamma_{4}) \{ \frac{1}{2} \sigma_{n}^{'}, 
\frac{1}{2} \lambda^{a}, \frac{1}{4}  \sigma_{n}^{'} \lambda^{a} \} 
\end{equation}
If we work with the set of generators $\frac{1}{2} \sigma_{n}, 
\frac{1}{2} \lambda^{a}, \frac{1}{4}  \sigma_{n} \lambda^{a}$ where 
$n = 1, 2, 3$ and $a = 1, \cdots ,8$ we must use the transformed
Hamiltonian  $H^{'}$.

We now turn to another transformation of the Hamiltonian that
singles out a direction and therefore is valid on the light cone.
For extremely relativistic quarks the expansion of $H$ in $m^{-2} \nabla^{2}$
or the velocity square is not a good one. In this case we can remove
transverse components of $p$, say $p_{\perp}$  ($p_{1}$ and $p_{2}$ if 
the direction singled out is the third axis) by a unitary transformation. This
is the Cini-Touschek$^{\cite{ct}}$ transformation rediscovered by Melosh$^{\cite{mel}}$.
It is given by

\begin{equation}
W_{\perp} =exp \{ \frac{i}{2} \tan^{-1} \frac{p_{\perp}}{m} \mbox{\boldmath 
$\gamma$}_{\perp} \cdot {\bf p}_{\perp} \} 
\end{equation}
where

\begin{equation}
 \mbox{\boldmath $\gamma$}_{\perp} \cdot {\bf p}_{\perp}= \gamma_{1} p_{1} +
\gamma_{2} p_{2}, ~~~~ p_{\perp} = \sqrt{p_{1}^{2} + p_{2}^{2}}           
\end{equation}

\begin{equation}
p_{n} = -i \partial_{n} 
\end{equation}
Then the transformed Hamiltonian takes the form

\begin{equation}
H^{''} = W_{\perp}^{-1} H W_{\perp} = \gamma_{5} \sigma_{3} p_{3} +
\gamma_{4} \sqrt{ m^{2} + p_{\perp}^{2}} 
\end{equation}
Note that $H^{''}$ no longer commutes with $\sigma_{n}$ and 
$\gamma_{4} \sigma_{n}$ but only with the subset

\begin{equation}
\frac{1}{2} \gamma_{4} \sigma_{1},~~\frac{1}{2} \gamma_{4} \sigma_{2}
~~~{\rm and}~~~\frac{1}{2} \sigma_{3}     \label{eq:hd}
\end{equation}
that also generates an $SU(2)$ group. Then

\begin{equation}
\frac{1}{2} \sigma_{1}^{''} = \frac{1}{2} W_{\perp}^{-1} \gamma_{4} \sigma_{1}
W_{\perp} ,~~~~
\frac{1}{2} \sigma_{2}^{''} = \frac{1}{2} W_{\perp}^{-1} \gamma_{4} \sigma_{2}
W_{\perp} \nonumber 
\end{equation}

\begin{equation}
\frac{1}{2} \sigma_{3}^{''} = \frac{1}{2} W_{\perp}^{-1} \sigma_{3} W_{\perp} 
\end{equation}
also generate rotations and the generators of the $SU(6)_{W}$ group

\begin{equation}
\frac{1}{2} \sigma_{n}^{''} ,~~~~~\frac{1}{2} \lambda^{a} ,~~~~~
\frac{1}{4} \sigma_{n}^{''} \lambda^{a}
\end{equation}
commute with the original Hamiltonian $H$.

The operators Eq.(\ref{eq:hd}) are the Stech$^{\cite{ste}}$ spin operators. They
were rediscovered and called $W$-spin operators by Lipkin and
Meshkov$^{\cite{lim}}$. We have seen that they are not conserved quantities,
but their Cini-Touschek transforms $\sigma_{n}^{''}$ are conserved in the free
quark limit. This property was discovered by Melosh$^{ \cite{mel}}$.

The importance of the spin operators $\sigma_{n}^{''}$ is that they also
occur in the symmetry of the vertex in which the singled out
direction is given by the momentum transfer.

The meaning of $\gamma_{4}$ can be elucidated by exhibiting the transformation
properties of the conserved quantities$^{ \cite{scpr}, \cite{gbit}}$. We distinguish
between C-odd and C-even conserved covariant spin tensors. The
C-even tensor is given by

\begin{equation}
\Omega_{\mu \nu} = - i \int H_{\mu \nu 0} d^{3}x
\end{equation}
where

\begin{equation}
H_{\mu \nu \lambda} = - \frac{1}{4} \bar{\psi} \sigma_{\mu \nu}
\overleftrightarrow{\partial}_{\lambda} \psi = -\frac{i}{2}
\epsilon_{\mu \nu \lambda \rho} \bar{\psi} \gamma_{5} \gamma_{\rho} \psi + 
~~{\rm additional~ terms}                 \label{eq:hx}
\end{equation}
is conserved:

\begin{equation}
\partial_{\lambda} H_{\mu \nu \lambda} = 0 
\end{equation}
so that the $\Omega_{\mu \nu}$ are time independent. These are related to the Wigner spin as

\begin{equation}
W^{-1} \Omega_{0i}  W = 0 \nonumber
\end{equation}

\begin{equation}
\frac{1}{2} \epsilon_{ijk} W^{-1} \Omega_{jk}  W = \Omega_{i} = \frac{1}{2} 
\int (a_{p}^{\dag} \sigma_{i} a_{p} + b_{p}^{\dag} \sigma_{i} b_{p})d^{3}p 
\end{equation}

Hence they can be interpreted as quark spin plus antiquark spin and
are associated with the $\sigma_{i}^{'}$.

Note that $\Omega_{i}$ generate $SU(2)$ but not the covariant quantities
$\Omega_{\mu \nu}$.

The $C$-odd conserved covariant tensor is

\begin{equation}
\Omega_{5 \mu} = - i \int H_{\mu 0} d^{3}x 
\end{equation}
where

\begin{equation}
H_{\mu \nu} = -  \frac{i}{4m} \bar{\psi} \gamma_{5} \gamma_{\mu} 
\overleftrightarrow{\partial}_{\nu} \psi = \frac{1}{2} 
\bar{\psi} \tilde{\sigma}_{\mu \nu} \psi + {\rm add.~ terms}     \label{eq:hy}
\end{equation}
and 

\begin{equation}
\partial_{\nu} H_{\mu \nu} = 0
\end{equation}

Making the unitary $W$ transformation we find the non-covariant
objects

\begin{equation}
\Omega_{i}^{'} = W^{-1} \Omega_{5i} W = \frac{1}{2} 
\int (a_{p}^{\dag} \sigma_{i} a_{p} - b_{p}^{\dag} \sigma_{i} b_{p})d^{3}p 
\end{equation}

\begin{equation}
W^{-1} \Omega_{50} W = O
\end{equation}
These conserved quantities are the quark spin minus the antiquark spin and correspond to 
$\gamma_{4} \sigma_{i}^{'}$. The $\frac{1}{2}  (\Omega_{i} \pm \Omega_{i}^{'})$ generate
the $SU(2) \times SU(2)$ group of quark and antiquark spin. They 
correspond to $\frac{1}{2} (1 \pm \gamma_{4}) \sigma_{i}^{'}$. These are 
sometimes called the $f$ spin and the $\bar{f}$ spin.

When we do the Cini-Touschek-Melosh transformation on $\Omega_{\mu \nu}$ and
$\Omega_{5 \mu }$, the $C$-odd operators $\Omega_{1}^{'}, \Omega_{2}^{'}$ and 
the $C$-even $\Omega_{3}$ commute with the Hamiltonian $H^{''}$.

If now we let $p_{3} \rightarrow \infty$ then

\begin{equation}
\Omega_{3} \rightarrow \frac{1}{2} \sigma_{3} \rightarrow \frac{1}{2}
\gamma_{5} 
\end{equation}
and the $C$-even subgroup of $SU(6)$ becomes the chiral $SU(3) \times SU(3)$
group whose generators commute with massless Dirac Hamiltonian,
so that

\begin{equation}
\lim_{p_{3} \rightarrow \infty} [SU(3) \times SU(3)]_{\sigma_{3}} = [SU(3) \times SU(3)]_{\gamma_{5}}
\end{equation}

To sum up, the free Dirac Hamiltonian in the Foldy-Wouthuysen
form admits a $SU(2N)_{q} \times SU(2N)_{\bar{q}}$ symmetry where $N$ 
corresponds to the dimension of the internal space and $q$, $\bar{q}$ 
refer respectively to quarks and antiquarks. If a longitudinal direction is 
singled out by an additional term in the Lagrangian, then the Dirac Hamiltonian
can be transformed by a Cini-Touschek transformation relative to the 
longitudinal direction and the Hamiltonian has a $SU(2N)_{W}$ invariance associated with the
subgroup of $SU(2N)_{q} \times SU(2N)_{\bar{q}}$ involving the $C$-odd 
transverse spin and the $C$-even longitudinal spin.

\section*{ Vertex Symmetry for Interacting Fermions}

     Consider the coupling of the fermion $\psi$ to a boson $\phi$. If
the virtual boson with momentum $q$ is emitted by the fermion $\psi (p)$
in momentum space so that the fermion wave function changes from  
$\psi (p)$ to $\psi (p')$, then $p - p'$ = $q$ is the momentum transfer.
$\psi (p)$ and  $\psi  (p')$ may refer to states with different internal quantum
numbers such as $u$ and $d$ quarks.  In the approximation that these
states are degenerate in mass the momentum transfer $q$ is space-like
and singles out a space direction.  Thus a Yukawa interaction term
will involve a direction in momentum space (the $q$ direction) which
can be taken as the direction singled out by the Cini-Touschek
transformation.

     Then the interacting Hamiltonian can have a non-local ($q$-dependent) $SU(6)_{W}$ symmetry, provided the free boson Lagrangian also exhibits an $SU(6)_{q} \times SU(6)_{\bar{q}}$ symmetry generated by C-even 
and C-odd spin operators.  Hence we need two kinds of bosons to achieve
vertex symmetry:  C-even bosons $\phi$   and C-odd bosons $\omega_{\mu}$.  Those
which have the same parity can be grouped in the same multiplet. 
Negative parity bosons will transform like the $S$-states of the ($\bar{q}
q$) system, so that we can choose multiplets of $J^{PC}$ bosons which
transform like $\psi \gamma_{5} \psi$ ($0^{-+})$ and $\bar{\psi} \gamma_{\mu}
\psi$ ($1^{--})$.

     Let us write such a Yukawa coupling in the case of no internal
symmetry

\begin{equation}
{\cal L} = {\cal L}_{0} + {\cal L}_{int}   \nonumber
\end{equation}

\begin{equation}
{\cal L}_{0} = {\cal L}_{0}(\psi) + {\cal L}_{0}(\phi) + {\cal L}_{0} (\omega_{\mu}) \nonumber
\end{equation}

\begin{equation}
{\cal L}_{int} = i g \bar{\psi} (\gamma_{5} \phi + \gamma_{\mu} \omega_{\mu}) \psi   \label{eq:ab}
\end{equation}
where ${\cal L}_{0}(\psi)$, ${\cal L}_{0}(\phi)$  and 
${\cal L}_{0}(\omega_{\mu})$   are respectively free lagrangians for the
fermion, the pseudoscalar and the vector bosons, with 
${\cal L}_{0}(\omega_{\mu})$ chosen so that the equations of motion also yield the subsidiary condition

\begin{equation}
\partial_{\mu} \omega^{\mu} = 0       \label{eq:ac}
\end{equation}
essential for associating $\omega_{\mu}$ with a $1^{--}$ object, and ${\cal L}_{int}$ is the simplest non-derivative Yukawa coupling for the fermion-boson system.

We could now find the Fourier transform of ${\cal L}$ by introducing the Fourier
transforms $\tilde{\bar{\psi}}(p^{'})$, $\tilde{\psi} (p)$ and
$\tilde{\phi} (q)$, $\tilde{\omega}_{\mu} (q)$, and then do a Cini-Touschek
transformation with respect to $q$.  Instead we shall consider a
simplified situation corresponding to quasi-free heavy fermions
that will justify the static limit.  Then

\begin{equation}
\xi_{\pm} = \frac{1}{2} (1 \pm \gamma_{4}) \psi
\end{equation}
correspond approximately to the particles and antiparticles. 
Writing $M$ for the large quark mass, we can consider an effective
interaction of the quarks with mesons ($\phi$ and $\omega_{\mu}$) 
that are in fact $q \bar{q}$ bound states bound by gluon exchange forces.  
If $\psi$ is quasi-free we can write

\begin{equation}
\bar{\psi} \gamma_{5} \psi = \frac{1}{2M} \partial_{\mu} 
(\bar{\psi} \gamma_{5} \gamma_{\mu} \psi)     \nonumber
\end{equation}

\begin{eqnarray}
i g \phi \bar{\psi} \gamma_{5} \psi & =& \frac{- i g}{2M}
\bar{\psi} \gamma_{5} \gamma_{\mu} \psi \partial_{\mu} \phi + {\rm total~~ divergence}
\nonumber   \\
    & =& \frac{- i g}{2M}
\bar{\psi} \gamma_{5} \gamma_{4} \psi \partial_{0} \phi -
\frac{ i g}{2M}
\bar{\psi} \gamma_{5} \mbox{\boldmath $\gamma$} \cdot (\mbox{\boldmath $\nabla$} \phi)  \psi 
\end{eqnarray}

The first term is of the order $\frac{\mu}{2M}$ if $\mu$ is the meson mass and
negligible in the static limit.  In the second term the operator
$\gamma_{4} \gamma_{5} \mbox{\boldmath $\gamma$} = - \mbox{\boldmath 
$\sigma$}$  
commutes with $\gamma_{4}$ and as a result does not mix particle states
with antiparticle states, so that putting $\xi_{+} = \xi$ (particle wave
function) we can make the replacement

\begin{equation}
i g \phi \bar{\psi} \gamma_{5} \psi \longrightarrow 
\frac{i g}{2M} \xi^{\dag} \mbox{\boldmath $\sigma$} \cdot (\mbox{\boldmath $\nabla$} \phi)  \xi 
\end{equation}
                
     For the vector meson $\omega_{0}$ is not independent of the independent
variables $\omega_{i}$ ($i = 1,2,3)$ and we can write
$i g \bar{\psi} \mbox{\boldmath $\gamma$} \cdot \mbox{\boldmath $\omega$}
 \psi = g \psi^{\dag} \gamma_{5}
\mbox{\boldmath $\sigma$} \cdot \mbox{\boldmath $\omega$} \psi$.  Expressing
small components in terms of the large components in the free-Dirac
approximation, and neglecting total divergence terms we can make the replacement

\begin{equation}
g \psi^{\dag} \gamma_{5} \mbox{\boldmath $\sigma$} \cdot \mbox{\boldmath 
$\omega$} \psi \longrightarrow
\frac{i g}{2M} \xi^{\dag}[(\mbox{\boldmath $\sigma$} \times \mbox{\boldmath $\nabla$}) \cdot 
\mbox{\boldmath $\omega$}] \xi +
\frac{i g}{2M} \xi^{\dag} (\mbox{\boldmath $\nabla$} \cdot 
\mbox{\boldmath $\omega$})  \xi 
\end{equation}
                                               
In momentum space $i \mbox{\boldmath $\nabla$}$ gives the virtual space-like momentum $k$ of
the mesons, so that if we call $\sigma_{\parallel}$ and $\sigma_{\perp}$ the longitudinal and
transverse spin operators we have

\begin{equation}
i \mbox{\boldmath $\sigma$} \cdot \mbox{\boldmath $\nabla$} = |{\bf k}| 
\frac{\mbox{\boldmath $\sigma$} \cdot {\bf k}}{|{\bf k}|}=
|{\bf k}| ~\sigma_{\parallel} 
\end{equation}

\begin{equation}
i \mbox{\boldmath $\sigma$} \times \mbox{\boldmath $\nabla$} = |{\bf k}| 
\frac{\mbox{\boldmath $\sigma$} \times {\bf k}}{|{\bf k}|}=
|{\bf k}| ~\mbox{\boldmath $\sigma$}_{\perp} 
\end{equation}
and the interaction term takes the approximate form

\begin{eqnarray}
{\cal L}_{int} &=& \frac{g |{\bf k}|}{2M} \psi^{\dag} (\sigma_{\parallel} \phi +
\gamma_{4} \mbox{\boldmath $\sigma$}_{\perp} \cdot \mbox{\boldmath $\omega$}_{\perp} +
\omega_{\parallel}) \psi    \nonumber   \\
   & & 
\sim \frac{g |{\bf k}|}{2M} \xi^{\dag} \omega_{\parallel} \xi +
\frac{g |{\bf k}|}{2M} \xi^{\dag} (\sigma_{\parallel} \phi +
\mbox{\boldmath $\sigma$}_{\perp} \cdot \mbox{\boldmath $\omega$}_{\perp}) \xi  
\end{eqnarray}                                               
with $\phi$   and $\mbox{\boldmath $\omega$}_{\perp}$ forming an $SU(2)_{W}$ triplet.  
$\omega_{\parallel}=\frac{{\bf k} \cdot \mbox{\boldmath $\omega$}}{|{\bf k}|}$ is 
a singlet.  If we interpret ${\cal L}_{int}$ in $x$ space, then

\begin{equation}
|{\bf k}| = \sqrt{- \nabla^{2}} 
\end{equation}

\begin{equation}
\sigma_{\parallel} = W^{\dag} (-i \mbox{\boldmath $\nabla$}) \sigma_{3} W (-i \mbox{\boldmath $\nabla$}) =
\frac{-i \mbox{\boldmath $\sigma$} \cdot \mbox{\boldmath $\nabla$}}{\sqrt{- \nabla^{2}}}   \nonumber
\end{equation}

\begin{equation}
{\sigma_{\perp}}^{(1)} = W^{\dag} (-i \mbox{\boldmath $\nabla$}) \sigma_{1} W (-i \mbox{\boldmath $\nabla$}) \nonumber
\end{equation}

\begin{equation}
{\sigma_{\perp}}^{(2)} = W^{\dag} (-i \mbox{\boldmath $\nabla$}) \sigma_{2} W (-i \mbox{\boldmath $\nabla$}) 
\end{equation}

\begin{equation}
W = \sqrt{\frac{M+\mbox{\boldmath $\sigma$} \cdot \mbox{\boldmath $\nabla$}}{(M^{2} - 
\nabla^{2})^{\frac{1}{2}}}}
\end{equation}
with
\begin{equation}
W^{\dag} (-i \mbox{\boldmath $\nabla$}) ~ W (-i \mbox{\boldmath $\nabla$}) = 1
\end{equation}
Both $\sigma_{\parallel}$ and $\mbox{\boldmath $\sigma$}_{\perp}$  generate $SU(2)_{W}$ and 
combine with an internal $SU(N)$ to
give $SU(2N)_{W}$.  Performing a Cini-Touschek-Melosh transormation in
the ${\bf k}$ direction for the kinetic term we see that the generators
of this $SU(2N)_{W}$ commute with the transformed Hamiltonian including
the effective approximate interaction term.

     Here it is important to note that the Yukawa coupling Eq.(\ref{eq:ab})
is invariant under the local symmetry group $SO(4,1)$ acting on the
five dimensional representation ($\phi$, $\omega_{\mu}$) of $SO(4,1)$.  
But this group does not leave ${\cal L}_{0}$ invariant since Eq.(\ref{eq:ac}) that is a consequence
of the choice of ${\cal L}_{0}$ is not covariant under $SO(4,1)$.  Only the non-local $SU(2)_{W}$ is the
symmetry of ${\cal L} = {\cal L}_{0} + {\cal L}_{int}$.
If we include the
internal symmetry group $SU(3)$, then we can write a Yukawa
interaction term that has $SU(6,6)$ invariance instead of $SO(4,1)$ in
our simplified example.  This is the group introduced by Salam et. al.  As is well known, $SU(6,6)$ cannot be a symmetry of the Lagrangian.  But as far as we can separate positive and negative
frequency parts of $\psi$ we have an approximate $SU(6)_{W}$ symmetry for
the whole Lagrangian.  In OCD where the quarks are quasi-free inside
the hadron, such an approximation is justifiable because of
asymptotic freedom.  In that sense we may say that the approximate
vertex symmetry mixing spin and internal degrees of freedom like
isospin is better understood within the QCD framework.  The
covariant generalization involves the spin current densities
Eq.(\ref{eq:hx}) and Eq.(\ref{eq:hy}).  The corresponding covariant spin operators 
$\Omega_{\mu \nu}$ and $\Omega_{5 \mu}$ that are conserved in the free quark approximation do not
lead to a close algebra in general.  However it can be shown$^{ \cite{gbit}}$  
that the matrix elements of the transverse part of $\Omega_{5 \mu}$
and the longitudinal part of $\Omega_{\mu \nu}$ between one particle 
states do lead to closure and to $SU(6)_{W}$ symmetry in a special frame.

     Since mesons are extended objects, a quark-meson coupling such
as Eq.(\ref{eq:ab}) cannot be a local interaction.  In general there will be
a form factor $F(q)$, where $q$ is the momentum transfer at the vertex.
Expanding the form factor in powers of $q$ is equivalent to writing
effective interactions involving higher derivatives of the meson
fields.  As an example we can write an effective Lagrangian with
pseudovector coupling for the pseudoscalar meson and a Pauli
coupling for the vector meson $\omega_{\mu}$, so that

\begin{equation}
{\cal L}_{int}^{'} = \frac{f}{2M} \bar{\psi} [i \gamma_{5} \gamma_{\mu} 
\partial_{\mu} \phi + \frac{1}{2} \sigma_{\mu \nu} (\partial_{\mu} \omega_{\nu}
- \partial_{\nu} \omega_{\mu})] \psi          \label{eq:hq}
\end{equation}
where $f$ is a new coupling constant.  We can also write 

\begin{equation}
{\cal L}_{int}^{'} = \frac{f}{2M} \psi^{\dag} [\gamma_{5} \partial_{0} \phi
+ \mbox{\boldmath $\sigma$} \cdot \mbox{\boldmath $\nabla$} \phi + i \gamma_{5} \gamma_{4} 
\mbox{\boldmath $\sigma$} \cdot
(\partial_{0} \mbox{\boldmath $\omega$} - \mbox{\boldmath $\nabla$} \omega_{0})+ \gamma_{4} 
(\mbox{\boldmath $\sigma$} \times \mbox{\boldmath $\nabla$}) \cdot \mbox{\boldmath $\omega$}] \psi 
\end{equation}
                                               
Taking the momentum ${\bf k}$ of the exchanged mesons to be space-like and
along the third (longitudinal) direction, we find

\begin{equation}
{\cal L}_{int}^{'}= \frac{f}{2M}  |{\bf k}| \psi^{\dag} (\sigma_{\parallel} 
\phi + \gamma_{4} \mbox{\boldmath $\sigma$}_{\perp} \cdot 
\mbox{\boldmath $\omega$}_{\perp}) \psi + O(k^{2}) 
\end{equation}
again exhibiting the W-spin invariance of the vertex.

\section*{Symmetry in the Potential Approximation}

     In the one boson exchange approximation between heavy quarks
we consider the case when the quarks emit two gluons that can turn
into a color singlet, C-even, $q \bar{q}$ bound state (effectively a
pseudoscalar meson $\phi$) or the alternative case when they emit
three gluons that go into a color-singlet, C-odd, $q \bar{q}$ bound state
(effectively a vector meson $\omega_{\mu}$).  Then the mesons $\phi$  and
$\omega$ will have
effective coupling constants ($g$ for non derivative and $f$ for
derivative couplings) arising from the expansion of their form
factors.  We can use these couplings also to describle the baryon
baryon potential due to meson exchange$^{ \cite{nag}}$. Indeed the baryons
host three quasi-free quarks each.  Following the usual quark
additivity assumption we shall let one quark from each baryon
interact at a time, while the other (spectator) quarks remain idle.
When the active quarks emit several gluons in a colored state, that
state cannot propagate due to color confinement.  The active quarks
can make contact only through emission of two or three gluon color
singlet states which, in turn can create mesons in intermediate
states.  Hence $g$ and $f$ will also be effective coupling constants
for the baryon-meson interactions and will enter in the calculation
of effective baryon baryon potentials due to meson exchange.  Thus,
starting from a scenario in QCD we are able to justify the
Moeller-Rosenfeld$^{\cite{moeller}}$ model in which baryons exchange vector and
pseudoscalar mesons with symmetrical coupling constants. 
Accordingly we consider in the following the effective quark-meson
vertex with no renormalizability restriction on the form of the
Yukawa interaction.
     The $\phi$  exchange between quarks will give for the Fourier
transform $v({\bf k})$ of the potential

\begin{equation}
v({\bf k}) = {\frac{g}{2M}}^{2} (\mbox{\boldmath $\sigma$}^{(1)} \cdot {\bf k})
(\mbox{\boldmath $\sigma$}^{(2)} \cdot {\bf k}) (|{\bf k}|^{2} + m^{2})^{-1} 
\end{equation}
where ${\bf k}$ is the space-like momentum transfer (in a frame with $k_{0} = 0$),
$m$ is the meson mass, $M$ the quark mass (that becomes the baryon mass
when we add the masses of the spectator quarks) and $f$ the
pseudoscalar coupling constant.  Its Fourier transform gives the
potential due to $\phi$, namely$^{ \cite{nag}}$

\begin{equation}
V^{(\phi)} \approx \frac{m}{4 \pi} g^{2} \frac{m^{2}}{4M^{2}}
[\frac{1}{3} \mbox{\boldmath $\sigma$}^{(1)} \cdot \mbox{\boldmath $\sigma$}^{(2)} \phi + S_{12} \chi]  \label{eq:hm}
\end{equation}
where $\phi$ and $\chi$ are functions of $x = mr$, and $S_{12}$ is the tensor operator

\begin{equation}
S_{12} = \frac{3(\mbox{\boldmath $\sigma$}^{(1)} \cdot {\bf r})
(\mbox{\boldmath $\sigma$}^{(2)} \cdot {\bf r})}{r^{2}} -
\mbox{\boldmath $\sigma$}^{(1)} \cdot \mbox{\boldmath $\sigma$}^{(2)}     
\end{equation}
with

\begin{equation}
r = |{\bf r}_{1} - {\bf r}_{2}| 
\end{equation}
and

\begin{equation}
\phi (x) = \frac{e^{-x}}{x}, ~~~~~~~ \chi (x) = (\frac{1}{3} + \frac{1}{x} +
\frac{1}{x^{2}}) \frac{e^{-x}}{x}               \label{eq:bx}
\end{equation}
Similary, the exchange of the vector meson $\omega_{\mu}$  gives in the same
approximation of neglecting recoil

\begin{equation}
V^{(\omega)} \approx \frac{m}{4 \pi} \{ g^{2} ( 1 + \frac{m^{2}}{8M^{2}}) \phi
+g^{2} \frac{m^{2}}{4M^{2}} [\frac{2}{3} \mbox{\boldmath $\sigma$}^{(1)} \cdot 
\mbox{\boldmath $\sigma$}^{(2)} \phi - S_{12} \chi] \}           \label{eq:bb}
\end{equation}
In this limit we can also put

\begin{equation}
f=g \frac{m}{2M} 
\end{equation}

In this approximation the total potential $V$ is

\begin{equation}
V \approx \frac{m}{4 \pi}  g^{2} \{ ( 1 + \frac{m^{2}}{8M^{2}}) \phi
+ \frac{m^{2}}{4M^{2}}  \mbox{\boldmath $\sigma$}^{(1)} \cdot 
\mbox{\boldmath $\sigma$}^{(2)} \phi \}
\end{equation}
Introducing the projection operator
\begin{equation}
P^{S} = \frac{1}{2} (1 + \mbox{\boldmath $\sigma$}^{(1)} \cdot \mbox{\boldmath 
$\sigma$}^{(2)}) , ~~~~~(P^{S})^{2} = P^{S}                        \label{eq:hp}
\end{equation}
with eigenvalue $1$ on spin singlet states, we have

\begin{equation}
V = \frac{m}{4 \pi}  g^{2} \{ ( 1 - \frac{m^{2}}{8M^{2}}) \phi
+ \frac{m^{2}}{2 M^{2}} P^{S} \phi \}       \label{eq:hn}
\end{equation}
Hence the potential is a superposition of two spin independent
potentials namely $\phi$ and $P^{S} \phi$.  The tensor terms $S_{12}$
in $V^{\phi}$ and $V^{\omega}$ are
spin dependent and highly singular at $x = 0$.  If $\phi$  and $\omega$ have the
same mass and the same coupling constant these terms cancel,
resulting in a smooth wave function for the solution of the
Schr\"{o}dinger equation with $V$ as potential.  Rarita and 
Schwinger$^{\cite{rs}}$ observed that even in the case
$m_{\phi} \neq m_{\omega}$ the potential is smooth
enough for the Schr\"{o}dinger equation to be soluble provided that   
$g_{\phi} = g_{\omega}$.

In order to generalize this model to include isospin we note
that diagrams considered so far violate the Zweig rule$^{\cite{zw}}$ according
to which disconnected quark diagrams are suppressed.  The $\phi$ and $\omega$ that are exchanged in these Zweig-forbidden diagrams are not only color singlets but also flavor singlets, i.e. they have zero
isospin or unitary spin.  Another effective interaction between
quarks will occur through the exchange of color singlet but flavor
carrying $q \bar{q}$ bound states, such that $q$ and $\bar{q}$ are associated 
with different flavor quantum numbers.  For example we can have

\begin{equation}
d+u \longrightarrow  d + (u+\bar{d}) + d \longrightarrow  d + u    \label{eq:ho}
\end{equation}
where the bracketed $u$ and $\bar{d}$ form an intermediate bound state 
($\pi^{+}$ or $\rho^{+}$) that is exchanged between a $u$ quark and a $d$ 
quark.  Again we can call $g$ the effective coupling constant 
($\bar{u} d \pi^{+}$) or ($\bar{u} d \rho^{+}$). 
If $m$ is the mass of the exchanged meson and $M$ the common mass of $u$
and $d$ assumed to be degenerate, the potentials resulting from $\pi^{+}$ and
$\rho^{+}$ exchange will have respectively the forms Eq.(\ref{eq:hm}) and 
Eq.({\ref{eq:bb}), so that for the equal coupling constant and equal mass case 
the total potential will have the form Eq.(\ref{eq:hn}).  Including all the
charged and neutral states, the total $u-d$ potential results from
the exchange of $0^{-}$ mesons $\eta$, $\pi^{i}$ and $1^{-}$ mesons
$\omega_{\mu}$ and $\rho_{\mu}^{i}$, where $i = 1,2,3$ are isospin labels.  
Denoting the degenerate $(u d)$ pair by $\psi (I = 1/2)$ we have the 
effective interaction term

\begin{equation}
{\cal L}_{int} = \frac{i}{2} g \bar{\psi} ( \gamma_{5} \eta +\gamma_{\mu} 
\omega_{\mu} + \gamma_{5} \mbox{\boldmath $\tau$} \cdot \mbox{\boldmath $\pi$} + \gamma_{\mu}
\mbox{\boldmath $\tau$} \cdot \mbox{\boldmath $\rho$}_{\mu}) \psi 
\end{equation}
                 
     Including the spin states, the quark doublet (particles) form
the $4$ dimensional representation of $SU(4)$ while the bound state $\eta$   
is an $SU(4)$ singlet. The $\mbox{\boldmath $\pi$}$, $\omega$ and $\mbox{\boldmath $\rho$}$ form the 
adjoint $15$ dimensional representation of $SU(4)$.  With respect to the 
$SU(4)_{W}$ symmetry of the vertex, $\omega_{\parallel}$ is the $SU(4)_{W}$
singlet, while the $SU(4)_{W}$ ($15$) consists of

\begin{equation}
(\eta, \mbox{\boldmath $\omega$}_{\perp}),~~~~~ \rho_{\parallel}^{i},~~~~~
(\pi^{i}, \mbox{\boldmath $\rho$}_{\perp}^{i}) 
\end{equation}

The exchange of these mesons leads to the potential of the form

\begin{equation}
\alpha \phi + \frac{1}{4} \beta (1+ \mbox{\boldmath $\sigma$}^{(1)} \cdot 
\mbox{\boldmath $\sigma$}^{(2)})
(1+ \mbox{\boldmath $\tau$}^{(1)} \cdot \mbox{\boldmath $\tau$}^{(2)}) \phi
\end{equation}
that commutes with the $SU(4)$ generators found from 
$\frac{1}{4} (1+ \mbox{\boldmath $\sigma$}^{(1)} \cdot \mbox{\boldmath 
$\sigma$}^{(2)})$ and 
$\frac{1}{2} (1+ \mbox{\boldmath $\tau$}^{(1)} \cdot \mbox{\boldmath $\tau$}^{(2)})$. 
This potential is therefore spin and isospin independent in the
limit of degenerate masses and equal coupling.

     If the mass degeneracy is lifted by allowing $\pi_{i}$ and $\rho^{i}$
to have different masses, a tensor force $S_{12}$ occurs in the potential 
between $u$ and $d$, hence, between the proton and the neutron, so that the
deuteron acquires an electric quadrupole moment.  The right sign
and magnitude are found$^{\cite{jn}}$ if the $\rho$ is three to four times heavier
than the pion, in accordance with experiment.

     The $SU(4)$ symmetry we have found is broken when we add to the
potential recoil terms associated with the one boson exchange. 
These are proportional to ${\bf L} \cdot {\bf S}$ (spin-orbit) and 
$Q_{12}$ (quadrupole) given by

\begin{equation}
{\bf L} = {\bf r} \times {\bf p},~~~~~~~ {\bf S} = \frac{1}{2} 
(\mbox{\boldmath $\sigma$}^{(1)} + \mbox{\boldmath $\sigma$}^{(2)}) \nonumber
\end{equation}
\begin{equation}
Q_{12} = \frac{1}{2} [(\mbox{\boldmath $\sigma$}^{(1)} \cdot {\bf L}) 
(\mbox{\boldmath $\sigma$}^{(2)} \cdot {\bf L})+ (\mbox{\boldmath 
$\sigma$}^{(2)} \cdot {\bf L}) (\mbox{\boldmath $\sigma$}^{(1)} \cdot {\bf L})] 
\end{equation}
                                                   
     Using $\phi$ and $\chi$ defined by Eq.(\ref{eq:bx}) we can write the 
additional terms associated with $\omega$ exchange as

\begin{equation}
V_{add}^{\omega} = -\frac{m}{4 \pi} g^{2} \frac{3 m^{2}}{2 M^{2}}
(\frac{1}{x}+ \frac{1}{x^{2}}) \phi {\bf L} \cdot {\bf S} +
\frac{m}{4 \pi} g^{2} \frac{3 m^{4}}{16 M^{4}} \frac{1}{x^{2}} \chi Q_{12} 
\end{equation}

     The exchange of the pseudoscalar $\phi$ contributes no ${\bf L} \cdot {\bf S}$ or $Q_{12}$  
terms that may cancel the highly singular additional terms due to $\omega$  
exchange.  A $Q_{12}$ term that can cancel the $Q_{12}$ term can only come
from the exchange of a scalar meson $\sigma$.  
Accordingly we can enlarge our scheme by adding a scalar field $\sigma$ with 
the Yukawa interaction

\begin{equation}
{\cal L}_{int}^{\sigma} = g \bar{\psi} \psi \sigma
\end{equation}
to the interaction term Eq.(\ref{eq:ac}).  Then we have an additional
potential due to $\sigma$  exchange that reads$^{\cite{nag}}$

\begin{equation}
V^{\sigma} = -\frac{m}{4 \pi} g^{2} \{ (1-\frac{m^{2}}{8 M^{2}}) \phi
+ \frac{m^{2}}{2 M^{2}}
(\frac{1}{x}+ \frac{1}{x^{2}}) \phi {\bf L} \cdot {\bf S} +
\frac{3 m^{4}}{16 M^{4}} \frac{1}{x^{2}} \chi Q_{12} \}
\end{equation}

     In the total potential

\begin{equation}
V = V^{\phi} + V^{\omega} + V^{\sigma}
\end{equation}
the $Q_{12}$ terms cancel, leaving

\begin{equation}
V = \frac{m}{4 \pi} g^{2} \frac{m^{2}}{2 M^{2}} \phi
\{ P^{S} - 4 (\frac{1}{x}+ \frac{1}{x^{2}}) {\bf L} \cdot {\bf S} \}    \label{eq:ia}
\end{equation}

     The term with $P^{S}$ (where $P^{S}$ is given by Eq.(\ref{eq:hp})) is spin
independent while the term in ${\bf L} \cdot {\bf S}$ breaks this symmetry.

\section*{ Chiral Symmetry in the Extended Moeller-Rosenfeld (M-R) Model.}

 The M-R model without isospin is based on $\omega$ and $\phi$ exchange. 
We have seen that, neglecting recoil, the model gives a spin-independent 
potential between the fermions.  Including recoil we
get a highly singular potential with spin dependent, spin orbit and
quadrupole terms.  The extended M-R model, with the inclusion of a
scalar $\sigma$-meson gives a spin independent potential broken only by
relativistic spin-orbit terms.  But now, for vanishing fermion
mass, the Lagrangian exhibits a higher symmetry, namely invariance
under $U(1) \times SU(2)_{W}$ where $U(1)$ which rotates $\phi$ and $\sigma$ 
is a chiral group.  It follows that the extended model has both $SU(2)_{W}$
and chiral invariance.  If we introduce isovector fields $\pi^{i}$   
(pseudoscalar), $\rho_{\mu}^{i}$ (vector), $\delta^{i}$ (scalar), the potential becomes

\begin{equation}
W = (1 + \tau_{i}^{(1)} \tau_{i}^{(2)}) V  
\end{equation}
where $V$ is given by Eq.(\ref{eq:ia}).  The new potential is both spin and
isospin independent except for spin dependent ${\bf L} \cdot {\bf S}$
terms.  The generators of $SU(4)_{W}$ are

\begin{equation}
{\bf S} =\frac{1}{2}(\mbox{\boldmath $\sigma$}^{(1)}+ 
\mbox{\boldmath $\sigma$}^{(2)}),~~~~~ T^{i} =
\frac{1}{2}(\tau_{i}^{(1)} + \tau_{i}^{(2)}),~~~~~ {\bf S}~T^{i} 
\end{equation}

     $SU(4)_{W}$ becomes $SU(6)_{W}$   if the isospin is replaced by unitary
spin.  But the Lagrangian also has an additional chiral symmetry in
the zero quark mass limit.  If the masses of the $u$ and $d$ quarks are
neglected we have an $SU(2)_{L} \times SU(2)_{R}$ invariance in chromodynamics.
For free quarks, combination with spin leads to $SU(4)_{L} \times SU(4)_{R}$  
invariance. The $q \bar{q}$ states belonging to the adjoint representation
will have both positive and negative parities.  Adding singlets $\sigma$  
and $\phi$ we consider the $32$ states

\begin{equation}
E_{n} ,~~~~~ \sigma ,~~~~~ \phi ,~~~~~ \omega_{n} ~~~~~ (n = 1, 2, 3) \nonumber
\end{equation}

\begin{equation}
A_{n}^{i} , ~~~~~ \delta^{i} ,~~~~~ \pi^{i} ,~~~~~ \rho_{n}^{i} ~~~~~
(i = 1, 2, 3)
\end{equation}
where $i$ is the isospin index and $n$ labels the independent spatial
components of mesons with spin $1$.  Here $\phi$, $\pi_{i}$, $\omega_{n}$,
$\rho_{n}^{i}$ have negative parity while the scalar mesons 
($\sigma$, $\delta_{i}$) and the pseudovector mesons
($E_{n}$, $A_{n}^{i}$) have positive parity.  The scalar mesons $\sigma$ and
$\delta_{i}$ are necessary to cancel the $Q_{12}$ terms due to $\omega$ and 
$\rho$ exchange respectively.  Thus the smoothing out of the wave functions for
bound states of light quarks (for which the recoil corrections are
important) enlarges the $SU(4)_{W}$ vertex symmetry that combines spin
and isospin to include also the chiral symmetry $SU(2)_{L} \times SU(2)_{R}$.

     As indicated before, using the additivity principle for the
quarks inside the baryons and letting one quark in each baryon
exchange a meson at a time, both the $SU(4)_{W}$ symmetry and the
chiral symmetry are reflected in the potential between two baryons.

     Note that the cancellation of the $Q_{12}$ terms requires the
degeneracy of $\sigma$ with $\omega_{n}$ and $\delta^{i}$ with $\rho_{n}^{i}$.  
This corresponds to an $O(4)$ subgroup of $SU(4) \times SU(4)$ which seems 
to be broken less than $SU(4)$.

\section*{Larger group structure}

Now $U(4,4)$ has $64$ generators, whereas $SU(4,4)$ has $63$. Writing

\begin{equation}
SU(4,4) \supset SU(4) \times SU(4) \supset SU(4) ~~~{\rm or}~~~SU(4)_{W}
\end{equation}
and

\begin{equation}
SU(4,4) \supset SO(4,4) \supset SO(4) \times SO(4) \supset SO(4).
\end{equation}
Then part of the ${\cal L}$ that is in $(1)$-rep is

\begin{equation}
\bar{\psi} \epsilon \psi = \psi_{L}^{\dag} \psi_{R} \epsilon + {\rm h.c.}
\end{equation}

In $(28)$-rep of $SO(4,4)$ part of ${\cal{L}}$ is

\begin{eqnarray}
&  & \bar{\psi} [ \sigma_{\mu \nu} \{ (\partial_{\mu} \omega_{\nu} -
\partial_{\nu} \omega_{\mu}) + i \gamma_{5} (\partial_{\mu} E_{\nu} -
\partial_{\nu} E_{\mu}) + \frac{\mbox{\boldmath $\tau$}}{2} (i \gamma_{5} 
\mbox{\boldmath $\pi$} +
\mbox{\boldmath $\delta$})] \psi +  \nonumber \\
&  & \bar{\psi} i \gamma_{5} \gamma_{\mu} (E_{\mu} + \partial_{\mu} 
\tilde{\eta}) \psi + \bar{\psi} i \gamma_{\mu} \frac{\mbox{\boldmath $\tau$}}{2} \psi
\mbox{\boldmath $\rho$}_{\mu} =       \nonumber   \\
&  & \psi_{L}^{\dag} \underline{\mbox{\boldmath $\sigma$}} \{ ( \underline{\mbox{\boldmath $\nabla$}} \times
\underline{\mbox{\boldmath $\omega$}} + i \underline{\mbox{\boldmath $\nabla$}} \times \underline{{\bf E}} ) 
+( \partial_{0} \underline{{\bf E}} - \underline{\mbox{\boldmath $\nabla$}} E_{0}) + i (
\partial_{0} \underline{\mbox{\boldmath $\omega$}} - \underline{\mbox{\boldmath $\nabla$}} \omega_{0})
\} \psi_{R} + {\rm h.c.}    \nonumber  \\
&  & +  \psi_{L}^{\dag} i \mbox{\boldmath $\tau$} \cdot \mbox{\boldmath $\pi$} \psi_{R} +
\psi_{L}^{\dag} i \mbox{\boldmath $\tau$} \cdot \mbox{\boldmath $\delta$} \psi_{R} + {\rm h.c.} \nonumber  \\
&  & +  ( \psi_{L}^{\dag}  \mbox{\boldmath $\tau$} \psi_{L} + \psi_{R}^{\dag}  
\mbox{\boldmath $\tau$} 
\psi_{R}) \mbox{\boldmath $\rho$}_{0} + ( \psi_{L}^{\dag} \sigma_{n} 
\mbox{\boldmath $\tau$} \psi_{L} -
\psi_{R}^{\dag} \sigma_{n} \mbox{\boldmath $\tau$} \psi_{R}) \mbox{\boldmath 
$\rho$}_{n} \nonumber  \\
&  & + (\psi_{L}^{\dag} \psi_{L} - \psi_{R}^{\dag} \psi_{R})(E_{0} + 
\partial_{0} \tilde{\eta}) 
+(\psi_{L}^{\dag} \sigma_{n} \psi_{L} + \psi_{R}^{\dag} \sigma_{n} \psi_{R}   
(E_{n} + \partial_{n} \tilde{\eta})
\end{eqnarray}
and in $(35)$ of $SO(4,4)$ it is

\begin{eqnarray}
&  & \bar{\psi} [ i \gamma_{5} \eta + \frac{1}{2} \sigma_{\mu \nu} 
\frac{\mbox{\boldmath $\tau$}}{2} \{ (\partial_{\mu} \mbox{\boldmath $\rho$}_{\nu} -
\partial_{\nu} \mbox{\boldmath $\rho$}_{\mu}) + i \gamma_{5} (\partial_{\mu} {\bf a}_{\nu}
-\partial_{\nu} {\bf a}_{\mu}) \} ] \psi   \nonumber  \\
&  & \bar{\psi} \{ i \gamma_{\mu} \omega_{\mu} + i \gamma_{5} \gamma_{\mu}
\frac{\mbox{\boldmath $\tau$}}{2} \cdot ({\bf a}_{\mu} + \partial_{\mu} 
\mbox{\boldmath $\pi$}) \} \psi    \nonumber   \\
&  & = \psi_{L}^{\dag} [ i \eta + \mbox{\boldmath $\tau$} \sigma_{n} \cdot 
\{ (\underline{\mbox{\boldmath $\nabla$}}
\times \underline{\mbox{\boldmath $\rho$}})_{n} + i(\partial_{0} 
\mbox{\boldmath $\rho$}_{n} -
\partial_{n} \mbox{\boldmath $\rho$}_{0}) \nonumber \\ & & + (\partial_{0} {\bf a}_{n} -
\partial_{n} {\bf a}_{0}) + i  (\underline{\mbox{\boldmath $\nabla$}}
\times \underline{{\bf a}})_{n} \} ] \psi_{R} + {\rm h.c.}   \nonumber  \\
&  & +  (\psi_{L}^{\dag} \psi_{L} + \psi_{R}^{\dag} \psi_{R}) \omega_{0} +
 (\psi_{L}^{\dag} \sigma_{n} \psi_{L} - \psi_{R}^{\dag} \sigma_{n} \psi_{R})
\omega_{n}   \nonumber  \\
&  & + (\psi_{L}^{\dag} \mbox{\boldmath $\tau$} \psi_{L} - \psi_{R}^{\dag} 
\mbox{\boldmath $\tau$} 
\psi_{R}) (a_{0} + \partial_{0} \mbox{\boldmath $\pi$} ) \nonumber \\ & & +
(\psi_{L}^{\dag} \sigma_{n} \mbox{\boldmath $\tau$} \psi_{L} + \psi_{R}^{\dag} \sigma_{n} 
\mbox{\boldmath $\tau$} \psi_{R}) (a_{n} + \partial_{n} \mbox{\boldmath $\pi$} )
\end{eqnarray}

We now look at decomposition under $SU(4)_{L} \times SU(4)_{R}$ subgroup of
$SU(4,4)$. Parts containing
 
\begin{equation}
\psi_{L}^{\dag} \Omega \psi_{R} - \psi_{R}^{\dag} \Omega \psi_{L}   \nonumber
\end{equation}
terms have negative parity, while the parts with

\begin{equation}
\psi_{L}^{\dag} \Omega \psi_{R}+\psi_{R}^{\dag} \Omega \psi_{L}    \nonumber
\end{equation}
have positive parity. These lead to combined parts in ${\cal{L}}$

\begin{eqnarray}
&  & \psi_{L}^{\dag} \{ \epsilon + \sigma_{n} [ (\partial_{0} E_{n} -
\partial_{n} E_{0}) + ( \underline{\mbox{\boldmath $\nabla$}} \times \underline{\mbox{\boldmath 
$\omega$}})_{n}]
+ \mbox{\boldmath $\tau$} \cdot \mbox{\boldmath $\sigma$}   \nonumber   \\
&  & + \sigma_{n} \mbox{\boldmath $\tau$} \cdot [ (\partial_{0} {\bf a}_{n} -
\partial_{n} {\bf a}_{0}) + ( \underline{\mbox{\boldmath $\nabla$}} 
\times \underline{\mbox{\boldmath $\rho$}})_{n}] \} \psi_{R} + {\rm h.c.}  \nonumber   \\
&  & + \psi_{L}^{\dag} \{ i \mbox{\boldmath $\tau$} \cdot \mbox{\boldmath 
$\pi$} + i \sigma_{n}
(\partial_{0} \omega_{n} - \partial_{n} \omega_{0}) + i \sigma_{n}
( \underline{\mbox{\boldmath $\nabla$}} \times \underline{{\bf E}})_{n} + i \eta   \nonumber  \\
&  & +i \mbox{\boldmath $\tau$} \sigma_{n} \cdot 
(\partial_{0} \mbox{\boldmath $\rho$}_{n} - \partial_{n} \mbox{\boldmath 
$\rho$}_{0}) + i \mbox{\boldmath $\tau$}
\sigma_{n} \cdot ( \underline{\mbox{\boldmath $\nabla$}} \times \underline{{\bf a}})_{n}
\} \psi_{R} + {\rm h.c.}
\end{eqnarray}
where we put the positive parity parts in the first half, and the negative parity
parts in the second half of this equation. Now that $\tilde{\eta}$ is
a mixture of $\eta$ and $\eta^{'}$ we can now consider the $32$ states
($16$ negative and $16$ positive parity parts) which transform like
$(4, \bar{4}) + (\bar{4}, 4)$ under $SU(4)_{L} \times SU(4)_{R}$ with
$\partial_{\mu} E_{\mu} = 0$, $\partial_{\mu} \omega_{\mu} = 0$,
$\partial_{\mu} E_{\mu} = 0$, $\partial_{\mu} {\bf a}_{\mu} = 0$, and
$\partial_{\mu} \mbox{\boldmath $\rho$}_{\mu} = 0$. For time-like meson momentum,
$E_{0}$, $\omega_{0}$, ${\bf a}_{0}$, $\mbox{\boldmath $\rho$}_{0} = 0$, and for space-like
meson momentum in $O_{z}$ direction 
$E_{3}$, $\omega_{3}$, ${\bf a}_{3}$, $\mbox{\boldmath $\rho$}_{3} = 0$ so that
$E_{\parallel}$, $\omega_{\parallel}$, ${\bf a}_{\parallel}$, and
$\mbox{\boldmath $\rho$}_{\parallel}$ vanish. The $4 \times 4 = 1 + 15$ occur in
$\psi_{L}^{\dag} \psi_{L}$, $\psi_{R}^{\dag} \psi_{R}$ parts. $LL$, $RR$ parts
give the $31$ generators

\begin{equation}
(1, 1) +(15, 1) + (1, 15)
\end{equation}
put together with above $32$ giving the $63$ generators of $SU(4,4)$. One boson exchange of above
$(4, \bar{4}) + (\bar{4}, 4)$ gives on cancellation due to equality of $q-q-$meson coupling constant a potential of the form  

\begin{equation}
\frac{1}{2} (1 + \mbox{\boldmath $\tau$}^{(1)} \cdot \mbox{\boldmath $\tau$}^{(2)}) 
\frac{1}{2} (1 + \mbox{\boldmath $\sigma$}^{(1)} \cdot \mbox{\boldmath $\sigma$}^{(2)})~V(r)
\end{equation}
while the ${\bf L} \cdot {\bf S}$ term breaks the symmetry.

It is also possible to use non-compact form of $Sp(4) \supset SU(4,4)$. 

Making the $J^{PC}$ identifications $\tilde{\eta}$ ($0^{-+}$), 
$\mbox{\boldmath $\pi$}$
($0^{-+}$), $\omega_{\mu}$ ($1^{--}$), $\epsilon (f_{0})$ ($0^{++}$),
$\mbox{\boldmath $\delta$} ({\bf a}_{0})$ ($0^{++}$), $E_{\mu} (f_{1 \mu})$ ($1^{+-}$),
${\bf A}_{\mu} ({\bf b}_{\mu})$ ($1^{+-}$), under $SU(2)_{L} \times SU(2)_{R}$
exchanging particles $\tilde{\eta}$ with $\mbox{\boldmath $\delta$} ({\bf a}_{0})$, 
$\mbox{\boldmath $\pi$}$  with $\epsilon (f_{0})$, $\omega_{\mu}$ with
${\bf A}_{\mu} ({\bf b}_{\mu})$, and $\mbox{\boldmath $\rho$}_{\mu}$ with $E_{\mu} 
(f_{1 \mu})$ also leads to the same potential including relativistic
corrections (recoil).

\section*{Remarks on Symmetries Involving Spin and Internal Symmetries}

     We have seen that an approximate combination of spin and
internal degrees of freedom arises in effective Lagrangians where
the couplings of the $q \bar{q}$ bound states to quarks ($q$) or baryons 
($qqq$) are almost equal and the masses are almost degenerate.  The $SU(4)$,
$SU(6)$ or $SU(8)$ generators that close are not in general covariant. 
They are related to the covariant operators $\sigma_{n}$, 
$\sigma_{n} \lambda^{a}$ by a unitary transformation that is momentum 
dependent, hence non-local.  This
unitary transformation can be worked out for the free quark case. 
It serves as a model for the interacting case and works very well
in phenomenological applications$^{ \cite{hrw} \cite{cot} \cite{gkm}}$.  The 
success of this procedure is now partially understood by the asymptotic freedom of
quarks interacting through gluon exchange.  At short spatial
separation quarks are quasi free, so that their positive and
negative frequency parts can be separated and they can be subjected
to Foldy-Wouthuysen-Tani or Cini-Touschek-Melosh type
transformations.  The vertex then exhibits an $SU(6)_{W}$ type symmetry
for a quark emitting a meson with space-like momentum.  If the
meson momentum is time-like, the meson can only be coupled to a
quark-antiquark pair, each with a time-like momentum (in general
not on the mass shell).  In this case the effective interaction
Lagrangians of the form Eq.(\ref{eq:ab}) and Eq.(\ref{eq:hq}) have an $SU(2)$ (no
flavor), $SU(4)$ (isospin flavor) or $SU(6)$ (unitary spin flavor)
symmetry with the correspondences (in the $SU(4)$ case)

\begin{equation}
\tau^{i} \longleftrightarrow \pi^{i},~~~~~
\mbox{\boldmath $\sigma$} \longleftrightarrow \mbox{\boldmath $\omega$},~~~~~
\tau^{i} \mbox{\boldmath $\sigma$} \longleftrightarrow \mbox{\boldmath $\rho$}^{i}
\end{equation}
which give the usual $SU(4)$ classification for the bound states $q \bar{q}$. 
The spin dependence of the one-gluon exchange between $q$ and $\bar{q}$  
introduces a $\mbox{\boldmath $\sigma$}^{(1)} \cdot \mbox{\boldmath $\sigma$}^{(2)}$ term that splits
$\pi$ from $\omega$ and $\rho$ and leads to the mass formula for $SU(6)$

\begin{equation}
M = M_{0} + M_{1} Y + M_{2} [ I(I+1) - \frac{1}{4} Y^{2}] - M_{3} S(S+1) 
\end{equation}
that was proposed$^{\cite{gr}}$ in $1964$.  The $M_{0}$ part comes from the exact
$SU(6)$ limit valid for the spin independent confining potential$^{\cite{cgm}}$.
This spin independence is only seen in the lattice approximation in
QCD.  We have found a partial justification in the Moeller-Rosenfeld model 
with quarks exchanging almost degenerate $q\bar{q}$ bound
states.  The $S(S+1)$ part comes from short range one-gluon exchange
forces as shown by de R\'{u}jula, Georgi and Glashow$^{\cite{dgg}}$ who find the
value of $M_{3}$ to be proportional to $\alpha_{s}$, the QCD fine structure
constant and inversely proportional to the masses of the
constituent quarks.  Finally the $M_{1}$ and $M_{2}$ terms come from quark
mass difference.

     The spin independent confining force can be largely simulated
by an effective scalar field as shown by Schnitzer$^{\cite{schn}}$.

     A phenomenological bag model description of the hadron is
obtained by R. Friedberg and T. D. Lee$^{\cite{tdl}}$ if the phenomenological
scalar field (regarded as a function of invariants constructed out
of two-gluon fields and three-gluon fields) is assumed to have
different vacuum expectation values inside and outside the hadron. 
As shown by the same authors the large mass of the color and flavor
singlet can be understood by the contribution of the $2$-gluon
annihilation diagram for $q \bar{q}$.

     Once we have this phenomenological picture for hadrons we can
get non-trivial decay amplitude relations for mesons going into $2$
mesons or baryons decaying into baryons (by emission of mesons or
photons) by using the unitary transformation between the vertex
spin operators and the covariant spin operators.  The non-covariant
spin belongs to the $35$-dimensional adjoint representation of $SU(6)_{W}$. 
It is a function of the covariant spin operators of current algebra
and the momentum.  The momentum dependence can be taken into
account by consideration$^{\cite{bucs}}$ of the group $SU(6)_{W} \times SO(3)$ where the $SO(3)$ is associated with the orbital angular momentum exhibiting the momentum dependence.  The $SO(3)$ part can be described by the
quantum number $L = 0, 1, 2$, etc.. We have seen that in the
infinite momentum frame $\gamma_{5}$ converges to the longitudinal spin
$\sigma_{\parallel}$ or $\sigma_{3}$ (with $O_{2}$ being chosen as the 
longitudinal direction).  The pion operator $\gamma_{5} \mbox{\boldmath $\tau$}$ in 
current algebra becomes $\sigma_{3} \mbox{\boldmath $\tau$}$.  The conserved spin is
obtained by a unitary transformation of this operator.  It will
belong in general to a superposition of ($35$) $L = 0, L = 1, L = 2$,
etc..  For a given matrix element of the pion operator between two
states (meson states for pionic decay of mesons, baryonic state for
pionic decays of baryons) only a limited number of these
representations ($35, L$) will contribute.  That will immediately
give non-trivial relations between meson-meson-pion amplitudes or
between baryon-baryon-pion amplitudes. $\gamma$-decays or vector meson
decays are handled in similar fashion.  The details of this
analysis can be found in the reviews of Meshkov$^{\cite{mesh}}$ for meson
decays, Hey$^{\cite{hey}}$ for baryon decays or in the numerous original
articles$^{\cite{hrw} \cite{cot} \cite{gkm}}$ exploring the unitary 
transformation between the $SU(6)$ operators of the non-covariant 
constituent quarks and the $SU(6)$ operators of current algebra generated 
by covariant current quarks. 
More model dependent and detailed descriptions of the bound states
and decay processes are obtained in a semi-classical field theory
of quarks bound by a confining (harmonic or linear) potential that
is from the outset taken to be spin independent$^{\cite{cgm}}$.
 
     The chiral structure is related to the approximate zero mass
of the $u$ and $d$ quarks and the negligibly small mass of the pion$^{\cite{cp}}$. 
In quark phenomenology the $u$ and $d$ quark masses are usually taken
to be one third of the proton or neutron masses.  On the other
hand, if they are evaluated from the deviations from exact chiral
symmetry in current algebra, they are found to be of the order of
a few MeV.  The two pictures can be reconciled by putting massless
quarks in a bag (resulting from an effective scalar field) of
radius $R$, assuming that the sphere of radius $R$ is the boundary of
regions where the scalar field assumes different vacuum expectation
values.  Then, quarks behave as if they had a mass proportional to
$R^{-1}$ which can be adjusted to one third of the proton mass.  For an
attempt to understand the zero pion mass within this
phenomenological bag model based on QCD, see Friedberg and Lee's article$^{\cite{tdl}}$.

\end{document}